\documentclass[10pt,conference]{IEEEtran}

\usepackage{amssymb}

\usepackage{amsmath}

\newcommand{\myem}[1]{{\em #1}}
\newcommand{\F}{{\mathbb F}}
\newcommand{\refeq}[1]{(\ref{#1})}

\newtheorem{proposition}{Proposition}
\newtheorem{corollary}{Corollary}
\newtheorem{theorem}{Theorem}
\newtheorem{definition}{Definition}
\usepackage[latin1]{inputenc}

\begin{document}

\title{On formulas for decoding   binary cyclic codes}

\author{
\authorblockN{Daniel Augot}
\authorblockA{INRIA-Rocquencourt\\
Domaine de Voluceau\\
Le Chesnay, FRANCE}
\and
\authorblockN{Magali Bardet}
\authorblockA{Laboratoire LITIS\\ Université de Rouen }
\and
\authorblockN{Jean-Charles Faugère}
\authorblockA{
INRIA Rocquencourt, Salsa project\\
Université Pierre et Marie Curie-Paris 6\\
UMR 7606, LIP6
}
}

\maketitle

\begin{abstract}
  We adress the problem of the algebraic decoding of any cyclic code
  up to the true minimum distance. For this, we use the classical
  formulation of the problem, which is to find the error locator
  polynomial in terms of the syndroms of the received word. This is
  usually done with the Berlekamp-Massey algorithm in the case of BCH
  codes and related codes, but for the general case, there is no
  generic algorithm to decode cyclic codes.  Even in the case of the
  quadratic residue codes, which are good codes with a very strong
  algebraic structure, there is no available general decoding
  algorithm.

  For this particular case of quadratic residue codes, several authors
  have worked out, by hand, formulas for the coefficients of the
  locator polynomial in terms of the syndroms, using the Newton
  identities. This work has to be done for each particular quadratic
  residue code, and is more and more difficult as the length is
  growing. Furthermore, it is error-prone.

  We propose to automate these computations, using elimination theory
  and Gröbner bases. We prove that, by computing appropriate Gröbner
  bases, one automatically recovers formulas for the coefficients of the
  locator polynomial, in terms of the syndroms.
\end{abstract}


\section{Introduction}
There is a longstanding problem of efficiently decoding binary
quadratic residue codes. For each prime number $l$ such that 2 is a
quadratic residue modulo $l$, there exists essentially one such code.
It is a cyclic code of length $l$, whose defining set if the set of
the quaratic residue modulo $l$. It is proven that the minimu distance
of these codes is at least $\lfloor\sqrt l\rfloor$ (the square-root
bound). But compiled tables show that the minimum distance of these
codes is much better than this bound, and it is an open question to
find or to estimate the minimum distance of these codes, althought
some progress has been achieved~\cite{Voloch:PREPRINT2004}.

Up to date, there is no general decoding algorithm for the whole class
of quadratic residue codes. Several efforts have been put up for
particular cases, that is to say for each particular length, mainly by
Chen, Truong, Reed, Helleseth and
others~\cite{Reed-Yin-Truong:IEEE_IT1990,Reed-Yin-Truong-Holmes:IEE_CDT1990,Reed-Truong-Chen-Yin:IEEE_IT1992,Chen-Reed-Truong:IEE_CDT1994,Lu-Wuu-Cheng-Lu:IJSC1995,He-Reed-Truong-Chen:IEEE_IT2001,Chang-Truong-Reed-Cheng-Lee:IEEE_COMM2003,Truong-Chang-Chen-Lee:IEEE_COMM2005},
for the lengths 31, 23, 41, 73, 47, 71, 79, 97, 103 and 113. All these
decoding algorithms are based on the Newton identities, which involve
the so-called \myem{error locator polynomial} and the \myem{syndroms}
of the received word. These Newton identites are to be written for
each particular length, and then to be worked out for isolating the
coefficients of the locator polynomial in terms of the syndroms, while
eliminating the \myem{unknown syndroms}, which appear in the Newton
identities. This elimination procedure is hand crafted by the authors.
So it is tedious, prone to errors, and the authors eventually fail to
find formulas for the coefficients of the locator polynomial.

A separate path of research has been to use the theory of Gröbner
bases for decoding {\em any} cyclic code. It was originated by
Cooper~\cite{Cooper:CCSP:1990,Cooper:ISIT1991,Cooper:EL1991},
althought the results were unproven. Cooper uses an algebraic system
of equations, closely related to the decoding problem, but different
from the Newton identities. These works only deal with BCH codes.
Later, these algebraic systems have been studied by Loustaunau and von
York~\cite{Loustaunau-York:AAECC1997}, Caboara and
Mora~\cite{Caboara-Mora:AAECC2002}, for any cyclic code, and they give
proofs of the statements by Cooper. In this vein of research, one
studies the ideal generated by the system of equations, and tries to
prove that the symbolic locator polynomial belongs to this ideal. Then
this polynomial can be found by the computation of a Gröbner with
respect to a relevant ordering on the monomials.

Another system defined by the Newton identities has been considered by
Chen, Helleseth, Reed and
Truong~\cite{Chen-Reed-Helleseth-Truong:IEEE_IT1994b} (see
also~\cite{deBoer-Pellikaan:STOCAa,deBoer-Pellikaan:STOCAb}). In that
case, the aim is to prove that the ideal generated by the Newton
identities contains, for each coefficient $\sigma_i$ of the locator
polynomial, a polynomial of whose leading monomial is of degree one in
$\sigma_i$, and that this polynomial does not involve the unkown
syndroms.

\section{Our contribution}

We have already discussed the use of Gröbner bases for decoding cyclic
codes~\cite{Augot-Bardet-Faugere:ISIT2003} with a system different
from the Newton identities. At that time, we discussed the computation
of  Gröbner basss online: for each received word, one
computes the syndroms, and subsitutes them into an algebraic system of
equations. Then the computation of the Gröbner basis gives the
coefficients of the locator polynomials, which are sought for.

In this work, we discuss the idea of precomputing the Gröbner basis of
a system in which the syndroms are left as indeterminates. Then we
show that this Gröbner basis leads to formulas for the coefficients of
the locator polynomial. This is called \myem{one-step decoding}.

Still, there is the problem that these formulas for the coefficients
$\sigma_i$'s of the locator polynomial are of the form
$p_i\sigma_i+q_i=0$, where $p_i,q_i$ involve only the syndroms. Thus
finding $\sigma_i$ can be done as follows
\[
\sigma_i=\frac{q_i}{p_i},
\]
which may lead to a division by zero, when the actual values of the
syndromes are substituted into $p_i$.

Our second contribution is to introduce a new ideal, which contains
formulas of the form $\sigma_i+q_i=0$. Thus finding the $\sigma_i$'s
do not involve any division after substitution.

\section{Definitions}
We consider only binary cyclic codes. Let $n$ be the length, which is
odd, and $\alpha$ be a primitive $n$-th root of unity in some
extension $\F_{2^m}$ of $\F_2$. To each binary word
$c=(c_0,\dots,c_{n-1})$ of length $n$, is associated the polynomial
$c_0+c_1X+\dots+c_{n-1}X^{n-1}$. The Fourier Transform of $c$ is the
vector $S=(S_0,\dots,S_{n-1})$, with $S_i=c(\alpha^i)$. A cyclic code
is built by considering a \myem{defining set}
$Q=\{i_1,\dots,i_l\}\subset \{0,1,\dots,n-1\}$. The cyclic code $C$ of
defining set $Q$ is then the set of words whose Fourier Transform
satisfies
\[
S_{i_1}=\dots=S_{i_l}=0.
\]

Let $y\in \F_2^n$ the received word, to be decoded. As usual, we
write $y=c+e$, where $c$ is the codeword, and $e$ is the error. We
compute the Fourier Transform $S$ of $y$, and for $i\in Q$, we have:
\[
S_i=y(\alpha^i)=c(\alpha^i+e(\alpha^i)=e(\alpha^i),\quad i\in Q,
\]
since $c\in C$. The $S_i$'s, $i\in Q$ are called the \myem{syndroms}
of $e$, and the $S_j$'s, $j\not\in Q$ are the \myem{unknown syndroms}.
The decoding problem is to find $e$ given the syndroms $S_i$'s, $i\in
Q$, under the constraint that the weight of $e$ is bounded by
$t=\lfloor\frac{d-1}2\rfloor$, where $d$ is the minimum distance of
$C$, and the \myem{decoding radius} of $C$.

\section{The Newton's identities}
Let the error $e$ be of weight $w$, and let $u_1,\dots,u_w$ the
indices of the non zero coordinates of $e$.  These indices are encoded
in the \myem{locator polynomial} $\sigma(Z)$, defined as follows:
\[
\sigma(Z)=\prod_{i=1}^w(1-\alpha^{u_i}Z)=\sum_{i=0}^w\sigma_iZ^i,
\]
where $\sigma_1,\dots,\sigma_w$ are the \myem{elementary symmetric
  functions} of $\alpha^{u_1},\dots,\alpha^{u_l}$, which are called
the \myem{locators} of $e$. We note by $Z_1,\dots,Z_w$ the locators of
$e$. Finding $e$ is equivalent to finding $\sigma(Z)$, and the problem is
considered to be solved when $\sigma(Z)$ is found, thanks to the Chien
search~\cite{Chien:IEEE_IT1964}.

The Newton identities relate the elementary symmetric functions of the
locators of $e$ to the coefficients of the Fourier Transform of $e$.
They have the following form (see~\cite{Sloane-MacWilliams:TTOECC1983}):
\begin{equation}\label{eq:eq_i}
\begin{split}
\left\{\begin{array}{l}
  S_i+\displaystyle \sum_{j=1}^{i-1}\sigma_jS_{i-j}+i\sigma_i=0,\quad i\leq w,\\
  S_i+\displaystyle\sum_{j=1}^w\sigma_jS_{i-j}=0,\quad w < i \leq n+w .
  \end{array}\right.
\end{split}
\end{equation}
Note that the indices of the $S_i$ are cyclic, i.e.\ $S_{i+n}=S_i$.
In these equations, there are the $\sigma_i$'s, that we are looking
for, the $S_i$, $i\in Q$, and the $S_i$'s, $i\not\in Q$, that we try
to eliminate.  Our objective is to find an expression of the
$\sigma_i$'s in terms of the $S_i$'s, $i\in Q$.

\section{Elimination theory}

We consider the ideal $I_{N,w}$, generated by the Newton identities:
\begin{equation}\label{eq:ideal_eq_i}
\begin{split}
I_{N,w}:\left\langle\begin{array}{l}
  S_i+\displaystyle \sum_{j=1}^{i-1}\sigma_jS_{i-j}+i\sigma_i,\quad i\leq w\\
  S_i+\displaystyle\sum_{j=1}^w\sigma_jS_{i-j},\quad n+w\geq i> w 
  \end{array}\right\rangle
\end{split}.
\end{equation}
Let us note by $\sigma$ the set of the variables
$\sigma_1,\dots,\sigma_w$, by $S_Q$ the set $\{S_i;i\in Q\}$, and
$S_N$ the set $\{S_i,i\not \in Q\}$. Then we have that $I_{N,w}$ is an
ideal in the polynomial algebra $\F_2[\sigma,S_Q,S_N]$.

A \myem{Gröbner basis} of an ideal $I$ is a particular set of
generators of $I$, which is well behaved with respect to various
operations: it enables to test equalities of ideals, to test ideal
membership and so on. Due to lack of space, we will not recall to
formal definition here, which can be found
in~\cite{Cox-Little-Oshea:IVAA1992}. We recall that this notion
depends on a monomial ordering: for each particular monomial ordering
there exists a corresponding Gröbner basis. Of utmost importance for
us are the following considerations~\cite{Cox-Little-Oshea:IVAA1992}.
\begin{definition}
Let $I\subset\F_2[x_1,\dots,x_m]$. Then the ideal
\[
I_k=I\cap \F_2[x_{k+1},\dots,x_m]
\]
is the $k$-th \myem{elimination ideal}. It is the set of all the
relations that can be obtained on $x_{k+1},\dots,x_m$, by elimination
of the $k$ first variables $x_1,\dots,x_k$.
\end{definition}
\begin{proposition}
  Let $I\subset\F_2[x_1,\dots,x_m]$ be an ideal and let $G$ be a
  Gröbner basis for the lexicographical ordering, with
  $x_1>\dots>x_n$. Then, the set
  \[
  G_k=G\cap \F_2[x_{k+1},\dots,x_m]
  \]
  is a Gröbner basis of the $k$-th elimination ideal
  $I_k=I\F_2[x_{k+1},\dots,x_m]$.
\end{proposition}
Thus it is sufficient to compute a single Gröbner $G$, and to retain
the relevant polynomials, to eliminate the unwanted variables. For the
problem of decoding, we get:
\begin{proposition}
  Let be given a monomial ordering such that the $S_i$'s, $i\not\in Q$
  are greater than the $S_i$'s, $i\in Q$, and the $\sigma_i$'s. Let
  $G$ be a Gröbner basis of $I_{N,w}$ for this ordering. Then
\[
G\cap\F_2[\sigma,S_Q]
\]
is a Gröbner basis of the elimation ideal $I_{N,w}\cap \F_2[\sigma,S_Q]$.
\end{proposition}

This means that, if we compute a Gröbner basis of $I_{N,w}$ for a
relevant ordering, we find a (finite) basis of all the relations
between the $\sigma_i$'s and the $S_i$'s, $i\in Q$. The problem is
that these relations may not be of degree one in the $\sigma_i$'s. Our
aim is to prove that there exists relations of the form
$p_i\sigma_i+q_i$ in this ideal, where $p_i,q_i\in\F_2[S_Q]$.

\section{The variety associated to the Newton identities}

First we have to study $V(I_{N,w})$ the \myem{ variety associated} to
the ideal $I_{N,w}$. It is the set of all $\sigma_i$'s, $S_i$'s, which
satisfy the Newton identities. Note that we consider this variety in
$\overline \F_2$, the algebraic closure of $\F_2$. We have the
following Theorem, which is an extension of the main result
of~\cite{Augot:FFA1996}.
\begin{theorem}
  Let $(\sigma,S)$ be in $V(I_{N,w})$, with
  $\sigma=(\sigma_1,\dots,\sigma_w)\in\overline \F_2^w$ and $
  S=(S_0,\dots,S_{n-1})\in\overline \F_2^n$. Let $e$ be the inverse
  Fourier Transform of $S$. Note that \myem{a priori} $e$ has coordinates in
  $\overline \F_2$. Then
  
  1. the weight of $e$ is less than $w$;

  2. $e$ has indeed coordinates in $\F_2$;

  3. if $\sigma(Z)$ is the polynomial 
  \[
  1+\sum_{i=1}^w\sigma_iZ^i,
  \]
  and if $\sigma_e(Z)$ is the locator polynomial of $e$, then there
  exists an integer $l$ and a polynomial $G(Z)$ such that
  \[
    \sigma(Z) = \sigma_e(Z) G(Z)^2 Z^l.
  \]
\end{theorem}
\begin{proof}
Ommitted due to lack of space.
\end{proof}
From the NullStellenSatz~\cite{Cox-Little-Oshea:IVAA1992}, we have:
\begin{corollary}
  Let $I_{N,w}\cap \F_2[S_Q,S_N]$ be the elimination ideal of the
  $\sigma_i$'s. If $I_{N,w}$ is radical, then $I_{N,w}\cap
  \F_2[S_Q,S_N]$ is the set of all the relations between the
  coefficients of the Fourier Transform of the binary words of weight
  less than $w$.  Furthermore, if we eliminate the $S_i$'s, $i\not\in
  Q$, then $I_{N,w}\cap \F_2[S_Q]$ is the set of all the relations
  betwen the syndroms of the words of weight less than $w\leq t$.
\end{corollary}

\begin{corollary}
  Let $S_{Q,e}$ be the set of syndroms of some word $e$. Let $T_w$ be
  a basis of $ I_{N,w}\cap \F_2[S_Q]$, then $e$ has weight $w\leq t$ if
  and only if
  \begin{equation}\label{eq:crit1}
  t(S_{Q,e})=0,\mbox{for all } t \in T_v,\text{ for all } v \leq w.
  \end{equation}
\end{corollary}

\section{Radical ideals}
In the above, we have stumbled on the difficulty on proving that
$I_{N,w}$ is a radical ideal. We believe it is, but we have not been
able to prove it. To overcome this difficulty, we consider the ideal
$I_{N,w}^0$, where we add the ``field equations'' to ensure that the
$\sigma_i$'s and the $S_i$'s belong to the field $\F_{2^m}$. It is the
ideal
\begin{equation}
I_{N,w}^0=I_{N,w}+\left\langle\begin{array}{l}
    S_i^{2^m}+S_i,i\in
  \{0,\dots,n-1\},\\
\sigma_i^{2^m}+\sigma_i, i\in\{1,\dots,w\}
\end{array}\right\rangle.
\end{equation}
Thanks to these field equations, the ideal $I_{N,w}^0$ is radical, and has
dimension zero (it has a finite number of solutions). It is a
consequence of \cite[Chap. 2, Prop. 2.7]{Cox-Little-OShea:UAG2005},
which implies that, if an ideal contains, for each variable, a
squarefree univariate polynomial  in this variable, then it is radical.

One can prove the following.
\begin{theorem}\label{theo:field_eq}
  For each binary word $e$ of weight $w$ less than $t$, for each
  $i\in\{1,\dots,w\}$, the ideal $I_{N,w}^0$ contains a polynomial
  \[
  p_i\sigma_i+q_i,\]

  with $p_i,q_i\in\F_2[S_Q]$ such that $p_i(S_{Q,e})\neq 0$, where
  $S_{Q,e}$ is the set of the  syndroms of $e$.
\end{theorem}
\begin{proof} Ommitted due to lack of space.
\end{proof}

Thus the decoding algorithm could be:
\begin{enumerate}
\item (precomputation) For each $w\in\{1,\dots,t\}$, compute a Gröbner
  basis $G_w$ of $I_{N,w}^0$, for an ordering such that the $S_i$, $i\not
  \in Q$, are greater than the $\sigma_i$'s which in turn are greater
  than the $S_i$'s, $i\in Q$;
\item (precomputation) from each Gröbner basis $G_w$, for each $i$,
  collect all the relations $p_i\sigma_i+q_i$, call $\Sigma_{w,i}$
  this set;
\item (precomputation) from each Gröbner basis $G_w$, collect the
  polynomials in $G_w\cap \F_2[S_Q]$, call $T_w$ this set of polynomials;
\item (online) for each received word $y$, compute the syndroms
  $S_{Q,y}=S_{Q,e}$, where $e$ is the error to be found;
\item (online) find the weight $w_e$ of $e$ using the
  criterion~\refeq{eq:crit1}.
\item (online) for each $i\in\{1,\dots,w_e\}$:
  \begin{enumerate}
  \item find the relation $p_i\sigma_i+q_i\in\Sigma_{w_e,i}$ such that
    $p_i(S_{Q_e})\neq 0$
    \item solve for $\sigma_i$:
\[
\sigma_i=\frac{p_i(S_{Q_e})}{q_i(S_{Q_e})}
\]
  \end{enumerate}
\end{enumerate}

There are two difficulties with this approach. First, the Gröbner
basis can contain many polynomials of the form $p_i\sigma_+q_i$,
$i\in\{1,\dots,w\}$, as we have observed on examples. Second, the
field equations of the type $\sigma_i^{2^m}+\sigma_i$, and
$S_i^{2^m}+S_i$ can be of large degree, even though the length of the
code is moderate. For instance, in the case of the quadratic residue
code of length 41, the splitting field is $\F_{2^{20}}=\F_{1048576}$.
This means that $I_{N,w}^0$ contains equations of degree more than one
million, and the computation of the Gröbner basis is intractable.

It is natural to try to remove the field equations, and to consider
the ideal $I_{N,w}$ without the field equations.

\section{An augmented ideal}
The difficulty, as mentionned above, is that we have not proven that
$I_{N,w}$ is a radical ideal, which is a necessary ingredient, among
others, to prove Theorem~\ref{theo:field_eq}. We will build an ideal
which contains $I_{N,w}$, which is radical, and which will contain
``nice'' formulas. First we introduce the ideal $I_\sigma$
corresponding to the definitions of the elementary symmetric
functions, and $I_S$ corresponding to the definition of the
coefficients of the Fourier Transform:
\[
    I_{\sigma}=\left\langle{\sigma_i-\sum_{1\leq j_1<\dots<j_i\leq w}
      Z_{j_1}\dots Z_{j_i};i\in\{ 1,\dots, w\}}\right\rangle;
\]
and
\[
    I_{S}=\left\langle{\begin{array}{ll}S_i-\sum_{j=1}^w Z_j^i,&i\in\{1,\dots,
        {n+w}\};\\S_{i+n}-S_i,&i\in\{ 1,\dots, w \}\end{array}}\right\rangle.
\]
Note this ideal belongs to the polynomial ring $\F_2[\sigma,S,
Z_1,\dots,Z_w]$. When we eliminate the  $Z_i's$,  we have the following
\begin{proposition}
\[
  (I_{S}+I_{\sigma})\cap \F_q[S,\sigma]=I_{N}
\]
\end{proposition}
\begin{proof}
Omitted due to lack of space.
\end{proof}
Let us introduce the following polynomial:
\[
\Delta(Z_1,\dots,Z_w)=Z_1\cdots Z_w\prod_{1\leq i < j \leq w} \left(Z_i-Z_j\right).
\]
This polynomial has the property that, if the weight $w_e$ of $e$ of
the error is less than $w$, then one can extend the locators
$Z_1,\dots,Z_{w_e}$ into $Z_1,\dots,Z_w$, in a way such that
$Z_1,\dots,Z_w$ are zeros of $\Delta$.  In other words, it captures,
in some sense, the property of being of weight strictly less than $w$.

We need the definition of a saturated ideal, with respect to a polynomial.
\begin{definition}
  Let $I\subset\F[x_1,\dots,x_n]$ be an ideal, and
  $f\in\F[x_1,\dots,x_n]$ be given. The saturated ideal of $I$ with
  respect to $f$, denoted $I:f^\infty$, is the ideal
  \begin{equation}
  I:f^\infty=\left\{g\in\F[x_1,\dots,x_n]: f^mg\in I \text{ for some
      }m>0\right\}
  \end{equation}
\end{definition}
One has that, under some restrictions, the variety associated to the
saturated ideal $I:f^\infty$, does not contain the zeros of $f$.
\begin{proposition}  
  Let $I=\langle f_1,\dots,f_s\rangle \subset\F[x_1,\dots,x_n]$ be an
  ideal and  $f\in\F[x_1,\dots,x_n]$ be given.  Let 
  $y$ be a new indeterminate. Consider
  \[
  \tilde I=\langle f_1,\dots,f_s,1-fy\rangle\subset \F[x_1,\dots,x_n,y],
  \]
then 
$I:f^\infty=\tilde I \cap\F[x_1,\dots,x_n]$.
\end{proposition}
Thus the saturated ideal can be computed by a Gröbner basis
computation and elimination. Now we introduce the saturated ideal
\begin{equation}
(I_\sigma+I_S):\Delta^\infty
\end{equation}
Then 
\begin{proposition}
The ideal 
\[
(I_\sigma+I_S):\Delta^\infty
\]
contains the polynomials
\[
\begin{array}{l}
Z_i^n+Z_i, i\in\{1,\dots,w\},\\
\sigma_i^{2^m}+\sigma_i,i\in\{1,\dots,w\},\\
S_i^{2^m}+S_i, i\in\{0,\dots,n-1\}.
\end{array}
\]
\end{proposition}
\begin{proof}
Ommitted due to lack of space.
\end{proof}
In particular, it is a radical ideal. Then, by elimination of the
$Z_i$'s, we have the ideal $I_{N,w}^\infty$:
\[
I_{N,w}^\infty=\left((I_\sigma+I_S):\Delta^\infty\right)\cap\F_2[\sigma,S]\supset I_{N,w}
\]
Note that a basis of $I_{N,w}^\infty$ can be computed by computing a
Gröbner basis of $I_S+I_\sigma+(1-y\Delta)$ for an ordering
eliminating $y$ and $Z_i$'s, and by retaining the polynomials in terms
of the $\sigma_i$'s and the $S_i$'s. Note also that $I_{N,w}^\infty$
is a radical ideal.

The variety associated to $I_{N,w}^\infty$ can be described as follows:
\begin{theorem}
  The variety $V(I_{N,w}^\infty)$ is exactly the set of the elementary
  symmetric functions and the elementary power-sum functions of the
  words of weight exactly $w$.
\end{theorem}
\begin{proof}
Ommitted due to lack of space.
\end{proof}
In particular,  we  have:
\begin{corollary}
  Let $S_{Q,e}$ be the set of syndroms of some word $e$. Let $T_w$ be
  a basis of $ I_{N,w}^\infty\cap \F_2[S_Q]$, then $e$ has weight $w$ if
  and only if
  \begin{equation}\label{eq:crit2}
  t(S_{Q,e})=0,\mbox{for all } t \in T_w.
  \end{equation}
\end{corollary}

Armed with this Theorem, and with the radicality of $I_{N,w}^\infty$, we
can prove:
\begin{theorem}\label{theo:nozerodiv}
  For each $i\in\{1,\dots,w\}$, $I_{N,w}^\infty$ contains a polynomial
  of the form $\sigma_i+q_i$, with $q_i\in\F_2[S_Q]$.
\end{theorem}
Note that this polynomial will appear in a Gröbner basis of
$I_{N,w}^\infty$, computed  as above.

The algorithm for
decoding is
\begin{enumerate}
\item (precomputation) For each $w\in\{1,\dots,t\}$, compute a Gröbner basis $G_w$ of
  $I_{N,w}^\infty$, written for the weight $w$;
\item (precomputation) From each $G_w$, for each $i$, pick the
  polynomial $q_{i,w}$ which appears in the polynomial
  $\sigma_i+q_{i,w}$ in Theorem~\ref{theo:nozerodiv}.
\item (precomputation) From each $G_w$, pick all the polynomials in
  $G_w\cap\F_2[S_Q]$, call $T_w$ this set of polynomials;
\item (online) for each received word $y$, compute the syndroms
  $S_{Q,y}=S_{Q,e}$, where $e$ is the error to be found;
\item (online) for each possible weight $w$ of the error, find the
  weight $w_e$ of the error using the criterion~\refeq{eq:crit2}.
\item (online) compute $\sigma_i=q_{i,w_e}(S_{Q,e})$.
\end{enumerate}

Thus we have removed the problem of the field equations, and the
problem of the division by zero.

\section{Conclusion}
For the decoding of any cyclic code, up to the true minimum distance,
we have shown how to find relations of degree one for the coefficients
of the locator polynomials, in terms of the syndroms. These relations
can be computed from the Newton identities. Then we have introduced an
ideal containing the ideal generated by the Newton identities, which
give formulas for the coefficient of the locator polynomial, with no
leading terms (and thus avoiding the problem of dividing by zero).

\IEEEtriggeratref{16}


\begin{thebibliography}{10}
\providecommand{\url}[1]{#1}
\csname url@rmstyle\endcsname
\providecommand{\newblock}{\relax}
\providecommand{\bibinfo}[2]{#2}
\providecommand\BIBentrySTDinterwordspacing{\spaceskip=0pt\relax}
\providecommand\BIBentryALTinterwordstretchfactor{4}
\providecommand\BIBentryALTinterwordspacing{\spaceskip=\fontdimen2\font plus
\BIBentryALTinterwordstretchfactor\fontdimen3\font minus
  \fontdimen4\font\relax}
\providecommand\BIBforeignlanguage[2]{{%
\expandafter\ifx\csname l@#1\endcsname\relax
\typeout{** WARNING: IEEEtran.bst: No hyphenation pattern has been}%
\typeout{** loaded for the language `#1'. Using the pattern for}%
\typeout{** the default language instead.}%
\else
\language=\csname l@#1\endcsname
\fi
#2}}

\bibitem{Voloch:PREPRINT2004}
\BIBentryALTinterwordspacing
F.~Voloch, ``Asymptotics of the minimal distance of quadratic residue codes,''
  2004. [Online]. Available:
  \url{\url{http://www.ma.utexas.edu/users/voloch/preprint.html}}
\BIBentrySTDinterwordspacing

\bibitem{Reed-Yin-Truong:IEEE_IT1990}
\BIBentryALTinterwordspacing
I.~S. Reed, X.~Yin, and T.-K. Truong, ``Algebraic decoding of the (32, 16, 8)
  quadratic residue code,'' \emph{IEEE Transactions on Information Theory},
  vol.~36, no.~4, pp. 876--880, 1990. [Online]. Available:
  \url{http://ieeexplore.ieee.org/xpls/abs\_all.jsp?arnumber=53750}
\BIBentrySTDinterwordspacing

\bibitem{Reed-Yin-Truong-Holmes:IEE_CDT1990}
\BIBentryALTinterwordspacing
I.~S. Reed, X.~Yin, T.-K. Truong, and J.~K. Holmes, ``Decoding the (24,12,8)
  golay code,'' \emph{Computers and Digital Techniques, IEE Proceedings-}, vol.
  137, no.~3, pp. 202--206, 1990. [Online]. Available:
  \url{http://ieeexplore.ieee.org/xpls/abs\_all.jsp?arnumber=50613}
\BIBentrySTDinterwordspacing

\bibitem{Reed-Truong-Chen-Yin:IEEE_IT1992}
\BIBentryALTinterwordspacing
I.~S. Reed, T.-K. Truong, X.~Chen, and X.~Yin, ``The algebraic decoding of the
  (41, 21, 9) quadratic residue code,'' \emph{IEEE Transactions on Information
  Theory}, vol.~38, no.~3, pp. 974--986, 1992. [Online]. Available:
  \url{http://ieeexplore.ieee.org/xpls/abs\_all.jsp?arnumber=135639}
\BIBentrySTDinterwordspacing

\bibitem{Chen-Reed-Truong:IEE_CDT1994}
\BIBentryALTinterwordspacing
X.~Chen, I.~S. Reed, and T.-K. Truong, ``Decoding the (73, 37, 13) quadratic
  residue code,'' \emph{IEE Proceedings on Computers and Digital Techniques},
  vol. 141, no.~5, pp. 253--258, 1994. [Online]. Available:
  \url{http://ieeexplore.ieee.org/xpls/abs\_all.jsp?arnumber=326788}
\BIBentrySTDinterwordspacing

\bibitem{Lu-Wuu-Cheng-Lu:IJSC1995}
E.~H. Lu, H.~P. Wuu, Y.~C. Cheng, and P.~C. Lu, ``Fast algorithms for decoding
  the (23,12) binary {G}olay code with four-error-correcting capability,''
  \emph{International Journal of Systems Science}, vol.~26, no.~4, pp.
  937--945, 1995.

\bibitem{He-Reed-Truong-Chen:IEEE_IT2001}
\BIBentryALTinterwordspacing
R.~He, I.~S. Reed, T.-K. Truong, and X.~Chen, ``Decoding the (47,24,11)
  quadratic residue code,'' \emph{IEEE Transactions on Information Theory},
  vol.~47, no.~3, pp. 1181--1186, 2001. [Online]. Available:
  \url{http://ieeexplore.ieee.org/xpls/abs\_all.jsp?arnumber=915677}
\BIBentrySTDinterwordspacing

\bibitem{Chang-Truong-Reed-Cheng-Lee:IEEE_COMM2003}
\BIBentryALTinterwordspacing
Y.~Chang, T.-K. Truong, I.~S. Reed, H.~Y. Cheng, and C.~D. Lee, ``Algebraic
  decoding of (71, 36, 11), (79, 40, 15), and (97, 49, 15) quadratic residue
  codes,'' \emph{IEEE Transactions on Communications}, vol.~51, no.~9, pp.
  1463--1473, 2003. [Online]. Available:
  \url{http://ieeexplore.ieee.org/xpls/abs\_all.jsp?arnumber=1231644}
\BIBentrySTDinterwordspacing

\bibitem{Truong-Chang-Chen-Lee:IEEE_COMM2005}
\BIBentryALTinterwordspacing
T.-K. Truong, Y.~Chang, Y.-H. Chen, and C.~D. Lee, ``Algebraic decoding of
  (103, 52, 19) and (113, 57, 15) quadratic residue codes,'' \emph{IEEE
  Transactions on Communications}, vol.~53, no.~5, pp. 749--754, 2005.
  [Online]. Available:
  \url{http://ieeexplore.ieee.org/xpls/abs\_all.jsp?arnumber=1431115}
\BIBentrySTDinterwordspacing

\bibitem{Cooper:CCSP:1990}
A.~B. Cooper~III, ``Direct solution of bch syndrome equations,'' in
  \emph{Communications, Control, and Signal Processing}, E.~Arkian, Ed.\hskip
  1em plus 0.5em minus 0.4em\relax Elsevier, 1990, pp. 281--286.

\bibitem{Cooper:ISIT1991}
\BIBentryALTinterwordspacing
------, ``A one-step algorithm for finding {BCH} error locator polynomials,''
  in \emph{International Symposium ion Information Theory, ISIT 1991}, 1991,
  pp. 93--93. [Online]. Available:
  \url{\url{http://ieeexplore.ieee.org/xpls/abs\_all.jsp?arnumber=695149}}
\BIBentrySTDinterwordspacing

\bibitem{Cooper:EL1991}
\BIBentryALTinterwordspacing
------, ``Finding {BCH} error locator polynomials in one step,''
  \emph{Electronics Letters}, vol.~27, no.~22, pp. 2090--2091, 1991. [Online].
  Available: \url{http://ieeexplore.ieee.org/xpls/abs\_all.jsp?arnumber=133007}
\BIBentrySTDinterwordspacing

\bibitem{Loustaunau-York:AAECC1997}
\BIBentryALTinterwordspacing
P.~Loustaunau and E.~V. York, ``On the decoding of cyclic codes using {G}röbner
  bases,'' \emph{Applicable Algebra in Engineering, Communication and
  Computation}, vol.~8, no.~6, pp. 469--483, 1997. [Online]. Available:
  \url{http://dx.doi.org/10.1007/s002000050084}
\BIBentrySTDinterwordspacing

\bibitem{Caboara-Mora:AAECC2002}
\BIBentryALTinterwordspacing
M.~Caboara and T.~Mora, ``The {C}hen-{R}eed-{H}elleseth-{T}ruong decoding
  algorithm and the {G}ianni-{K}alkbrenner {G}röbner shape theorem,''
  \emph{Applicable Algebra in Engineering, Communication and Computing},
  vol.~13, no.~3, pp. 209--232, 2002. [Online]. Available:
  \url{http://dx.doi.org/10.1007/s002000200097}
\BIBentrySTDinterwordspacing

\bibitem{Chen-Reed-Helleseth-Truong:IEEE_IT1994b}
\BIBentryALTinterwordspacing
X.~Chen, I.~S. Reed, T.~Helleseth, and T.~K. Truong, ``General principles for
  the algebraic decoding of cyclic codes,'' \emph{IEEE Transactions on
  Information Theory}, vol.~40, no.~5, pp. 1661--1663, September 1994.
  [Online]. Available:
  \url{http://ieeexplore.ieee.org/xpls/abs\_all.jsp?arnumber=333886}
\BIBentrySTDinterwordspacing

\bibitem{deBoer-Pellikaan:STOCAa}
\BIBentryALTinterwordspacing
M.~de~Boer and R.~Pellikaan, \emph{Gröbner bases for codes}, ser. Algorithms
  and Computation in Mathematics.\hskip 1em plus 0.5em minus 0.4em\relax
  Springer, 1999, no.~4, pp. 237--259. [Online]. Available:
  \url{http://www.win.tue.nl/~ruudp/paper/34.pdf}
\BIBentrySTDinterwordspacing

\bibitem{deBoer-Pellikaan:STOCAb}
\BIBentryALTinterwordspacing
------, \emph{Gröbner bases for decoding}, ser. Algorithms and Computation in
  Mathematics.\hskip 1em plus 0.5em minus 0.4em\relax Springer, 1999, no.~4,
  pp. 260--275. [Online]. Available:
  \url{http://www.win.tue.nl/~ruudp/paper/35.pdf}
\BIBentrySTDinterwordspacing

\bibitem{Augot-Bardet-Faugere:ISIT2003}
\BIBentryALTinterwordspacing
D.~Augot, M.~Bardet, and J.-C. Faug\`ere, ``Efficient decoding of (binary)
  cyclic codes above the correction capacity of the code using {G}r\"obner
  bases,'' in \emph{Proceedings of the 2003 IEEE International Symposium on
  Information Theory}, 2003, pp. 362--362. [Online]. Available:
  \url{http://ieeexplore.ieee.org/xpls/abs\_all.jsp?arnumber=1228378}
\BIBentrySTDinterwordspacing

\bibitem{Chien:IEEE_IT1964}
\BIBentryALTinterwordspacing
R.~T. Chien, ``Cyclic decoding procedures for {B}ose-{C}haudhuri-{H}ocquenghem
  codes,'' \emph{IEEE Transactions on Information Theory}, vol.~10, no.~4, pp.
  357--363, 1964. [Online]. Available:
  \url{http://ieeexplore.ieee.org/xpls/abs\_all.jsp?arnumber=1053699}
\BIBentrySTDinterwordspacing

\bibitem{Sloane-MacWilliams:TTOECC1983}
F.~J. Macwilliams and N.~J.~A. Sloane, \emph{The Theory of Error-Correcting
  Codes}, ser. North-Holland Mathematical Library.\hskip 1em plus 0.5em minus
  0.4em\relax North Holland, January 1983.

\bibitem{Cox-Little-Oshea:IVAA1992}
D.~Cox, J.~Littel, and D.~O'Shea, \emph{Ideals, Varieties and
  Algorithms}.\hskip 1em plus 0.5em minus 0.4em\relax Springer, 1992.

\bibitem{Augot:FFA1996}
D.~Augot, ``Description of minimum weight codewords of cyclic codes by
  algebraic systems,'' \emph{Finite Fields Appl.}, vol.~2, no.~2, pp. 138--152,
  1996.

\bibitem{Cox-Little-OShea:UAG2005}
D.~A. Cox, J.~Little, and D.~O'Shea, \emph{Using Algebraic Geometry}, ser.
  Graduate Texts in Mathematics.\hskip 1em plus 0.5em minus 0.4em\relax
  Springer, March 2005.

\end{thebibliography}
\end{document}